\documentclass[10pt]{article}

\usepackage[superscript,sort]{cite}

\usepackage{ifthen,ifpdf}
\ifpdf
	\usepackage[pdftex]{hyperref}	\hypersetup{colorlinks=true,linkcolor=black,citecolor=black,urlcolor=blue,backref=page,bookmarks=true,breaklinks=true,plainpages=false }
	\usepackage[pdftex]{graphicx}
	\usepackage[usenames,pdftex]{color}
\else
	\usepackage[colorlinks=true,breaklinks=true,plainpages=false]{hyperref}
	\usepackage{graphicx}
	\usepackage[usenames]{color}
\fi

\usepackage{amsthm,amscd,amsxtra,amsfonts,amsmath,amssymb,multirow}
\usepackage{wrapfig}
\usepackage[footnotesize]{caption}
\usepackage[tiny,compact]{titlesec}
\usepackage[textwidth=0.8in,textsize=footnotesize]{todonotes}
\usepackage{algorithm,algorithmic,extarrows}

\begin{document}

\title{Multiscale virtual particle based elastic network model (MVP-ENM) for biomolecular normal mode analysis}

\author{
Kelin Xia$^{1,2}$ \footnote{ Address correspondences  to Kelin Xia. E-mail:xiakelin@ntu.edu.sg}\\
$^1$Division of Mathematical Sciences, School of Physical and Mathematical Sciences, \\
Nanyang Technological University, Singapore 637371\\
$^2$School of Biological Sciences \\
Nanyang Technological University, Singapore 637371\\
}

\date{\today}
\maketitle

\begin{abstract}
In this paper, a multiscale virtual particle based elastic network model (MVP-ENM) is proposed for biomolecular normal mode analysis. The multiscale virtual particle model is proposed for the discretization of biomolecular density data in different scales. Essentially, the model works as the coarse-graining of the biomolecular structure, so that a delicate balance between biomolecular geometric representation and computational cost can be achieved. To form ``connections" between these multiscale virtual particles, a new harmonic potential function, which considers the influence from both mass distributions and distance relations, is adopted between any two virtual particles. Unlike the previous ENMs that use a constant spring constant, a particle-dependent spring parameter is used in MVP-ENM. Two independent models, i.e., multiscale virtual particle based Gaussian network model (MVP-GNM) and multiscale virtual particle based anisotropic network model (MVP-ANM), are proposed. Even with a rather coarse grid and a low resolution, the MVP-GNM is able to predict the Debye-Waller factors (B-factors) with considerable good accuracy. Similar properties have also been observed in MVP-ANM. More importantly, in B-factor predictions, the mismatch between the predicted results and experimental ones is predominantly from higher fluctuation regions. Further, it is found that MVP-ANM can deliver a very consistent low-frequency eigenmodes in various scales. This demonstrates the great potential of MVP-ANM in the deformation analysis of low resolution data. With the multiscale rigidity function, the MVP-ENM can be applied to biomolecular data represented in density distribution and atomic coordinates. Further, the great advantage of my MVP-ENM model in computational cost has been demonstrated by using two poliovirus virus structures. Finally, the paper ends with a conclusion.
\end{abstract}

\newpage

{\setcounter{tocdepth}{5} \tableofcontents}

\newpage

\section{Introduction}
It is well known that the flexibility and deformational motions are of essential importance to biomolecular functions. Experimentally, biomolecular flexibility can be  directly or indirectly measured by X-ray crystallography, nuclear magnetic resonance (NMR) and single-molecule force experiments \cite{Dudko:2006}.  Debye-Waller factors, also known as  B-factors or temperature factors, are widely-used to characterize the uncertainty for each atom. Computationally, molecular dynamics (MD) \cite{McCammon:1977} has proved to be a powerful tool for the study of biomolecular conformational changes. However, MD tends to fall short when analyzing motions of macromolecules or biomolecular complexes at long time scale, due to computational limitations. And alternative approaches, including normal mode analysis (NMA)  \cite{Go:1983,Tasumi:1982,Brooks:1983,Levitt:1985}, graph theory \cite{Jacobs:2001} and elastic network model (ENM) \cite{Bahar:1997,Bahar:1998,Atilgan:2001,Hinsen:1998,Tama:2001,LiGH:2002,QCui:2010} become the main workhorses for biomolecular flexibility and collective motion analysis during the past two decades. The essential hypothesis of NMA is that lowest-frequency related vibrational normal modes captures the largest movements of a biomolecule and they are the ones that are functionally relevant\cite{Skjaven:2009,QCui:2010}. More specifically, a harmonic approximation of potential energy function near the energy-minimized configuration is used in NMA. Molecular motions are then decomposed into a series of vibrational modes. Interestingly, among all the vibrational normal modes, the overall protein deformational motions can be well represented with only a few low-frequency ones. Further, Tirion simplifies the potential functions in NMA into harmonic functions with a uniform force constance. This is the so-called elastic network model (ENM) \cite{Hinsen:1998}. ENM can dramatically reduce the computational time, largely due to two reasons. First, no energy minimization is required as the crystal structure is regarded as the energy minima state. Second, in coarse-grained (CG) ENM, harmonic potentials are only considered between $C_{\alpha}$ atoms within a small cutoff distance. Two most popular models from ENM are Gaussian network model (GNM)  \cite{Bahar:1997,Bahar:1998}  and anisotropic network model (ANM) \cite{Atilgan:2001}. ANM can be viewed as a coarse-grained ENM. In contrast to ANM, GNM can only be used to describe isotropic positional fluctuations. It can not be used to study conformational changes of biomolecule as ANM. However, GNM is much more efficient and accurate in Debye-Waller factors (B-factors) prediction than ANM \cite{LWYang:2008,JKPark:2013, Opron:2014}. In fact, GNM is about one order more efficient than most other flexibility approaches as  demonstrated by Yang et al. \cite{LWYang:2008}.

Two major challenges still exist for modeling of biomolecular flexibility and dynamic. The first one comes from extremely large biomolecular complexes. 
\cite{Keskin:2002,Tama:2003,YWang:2004,Rader:2005,Tama:2005,WZheng:2007}. The large degrees of freedom have disallowed traditional normal mode analysis, not to mention molecular dynamic simulations. Even for ENM at residual scale, a great number of structures are still computationally prohibitive. Due to the critical role of coarse-graining in tackling biomolecular complexity, in the past two decades, various coarse-graining approaches have been proposed, including rotation-translation of blocks (RTB)\cite{durand:1994new}, block normal mode (BNM) algorithm\cite{Tama:2000,LiGH:2002}, molecular symmetry based NMA\cite{van:2005normal}, NMA based fluctuation matching method (NMA-FM) \cite{chu:2005allostery,chu:2006coarse,xia:2012multiscale}, iterative matrix projection method\cite{na:2015bridging}, etc.
In RTB and BNM, a reduction of the biomolecular complexity is achieved by the coarse-graining of several residues into a residue block. Then, the low-frequency normal modes are approximated by rigid-body motions of the residual blocks. In molecular symmetry based NMA, the Cartesian basis in Hessian matrix are replaced by symmetry coordinate basis, which include only internal dihedral angles. In NMA-FM method, based on the potential of a fine-grained model either from molecular dynamics or elastic network model, a CG model is uniquely designed, so that physical properties, particularly fluctuations, predicted by the two representations can be matched very well. Usually, this method requires some effective internal coordinates to build its own potential. In the iterative matrix projection method, Hessian matrix from an all-atom representation is rewritten as a combination of CG part and residual part. An iterative projection is then used to generate a  coarse-grained Hessian matrix, which can still captures the dynamics of all-atom model very well. The iterative matrix projection method is derived from its previous spring-based NMA and simplified spring-based NMA  models\cite{na:2014bridging}.

The other challenge comes from the dynamic analysis of cryo-electron microscopy (Cryo-EM) data\cite{Wriggers:1999,Kuhlbrandt:2014}. Cryo-EM data are essentially biomolecular density distributions. For high resolution cryo-EM data, atomic information can be derived and biomolecular dynamics can be analyzed by MD, ENA and ANM. However, due to experimental limitations, lower resolution cryo-EM data have been frequently generated, particular for large biomolecular structures. Without detailed atomic information, the study of their flexibility and dynamic properties is by no means trivial. Quantized elastic deformation model(QEDM)\cite{Ming:2002describe,ming:2002domain,Tama:2002exploring,chacon2003mega} is the one the first methods proposed for collective motion analysis of cryo-EM data. The basic idea of QEDM is to treat a protein as an elastic object. 
Their conformational changes or collective global motions are largely determined by their global shapes and mass distributions\cite{lu2005:role,Ming:2002describe,Tama:2002exploring,tama2006symmetry}. In this way, vector quantization algorithm\cite{Gray:1984vector} is used to discretize a protein into a set of Voronoi cells\cite{wriggers:1998self,Wriggers:1999}, which characterize the geometric and density variation of the protein data. QEDM has been successfully used in B-factor and collective motion prediction. Another method for Cryo-EM data analysis is bend-twist-stretch (BTS) model, which has been proposed for an arbitrary coarse-graining of a biomolecular structure network by shifting the spatial complexity into potential function\cite{stember:2009bend}. The potential function has contributions from bond, bond angle and dihedral angles, and their force constants are derived from the continuum mechanics. In cryo-EM data analysis, BTS also employs the vector quantization for structural coarse-graining. And competitive Hebb rule\cite{martinetz:1993neural} is used to establish connection between adjacent pseudo-atoms, which are centroids of Voronoi cells. In this way, pseudo-bonds and pseudo-angles can also be defined, and a BTS model is constructed.

In this paper, a multiscale virtual particle based elastic network model (MVP-ENM) is introduced for biomolecular mode analysis. The basic ideas in MVP-ENM can be traced back to our flexibility-rigidity index  (FRI)  \cite{KLXia:2013d,Opron:2014}. FRI has been proved to be a highly accurate and efficient method for flexibility evaluation. Free from eigenvalue decomposition, the computational complexity of FRI is ${\cal O}(N^2)$ with $N$ the total number of $C_{\alpha}$ atoms. The fast FRI (fFRI) \cite{Opron:2014} employs cell lists algorithm \cite{Allen:1987} and is of ${\cal O}(N)$ complexity only. For an HIV virus capsid with 313 236 residues, fFRI is able to predict its B-factors in less than 30 seconds on a single-core  processor \cite{Opron:2014}. However, our FRI and fFRI can only be used in isotropic fluctuation analysis. Even though our anisotropic FRI (aFRI) can be used in normal mode prediction, it still requires eigenvalue decomposition of the Hessian matrix, thus it shares the same computational limitations as ANM. To solve this problem and further develop a new aFRI model for Cryo-EM data analysis, a virtual particle based aFRI model (VP-aFRI) is proposed\cite{Xia:2016review}. This model is based on the biomolecular density function, which is either generated from structure with atomic coordinates or directly obtained from Cryo-EM data. A virtual particle is defined as an element from the discretization of the domain of the density function. All virtual particles are correlated with each other and the spring constant between any two virtual particle is determined by their distance and mass distributions. Our VP-aFRI has been successfully used in collective motion analysis of Cryo-EM data\cite{Xia:2016review}.

In MVP-ENM, a multiscale virtual particle model is introduced and further combined with ENM to perform deformational motion analysis on large biomolecular structures. Similar to the previous VP-aFRI model\cite{Xia:2016review}, MVP-ENM is designed for biomolecular density data analysis. However, MVP-ENM employs a multiscale representation of the density data. To be more specific, for an atomic detailed structure, its bimolecular density distribution can be generated by a multiscale rigidity function, which uses a resolution parameter to control the generated density distribution into the scale of interest. Further, the discretization of molecular density profile, which is generated from the above rigidity function or directly obtained from Cryo-EM data,  can be done in various scales depending on the computational resources. The basic process in the MVP-ENM is following. Firstly, a molecular surface is extracted by using a suitable isovalue. Essentially, the surface is treated as the boundary of our computational domain. Secondly, the inside regions of the biomolecules, i.e., the regions enclosed by this surface, are discretized. The discretization can be done in many different ways. Cartesian grid as in the finite difference method, tetrahedron mesh as in the finite element method, Voronoi cell as in the tessalation method, etc, can all be employed in the domain decomposition. The individual elements from the discretization, i.e., Cartesian grid box, tetrahedron, Voronoi cell, etc, is defined as the virtual element or virtual particle. Thirdly, an elastic network is constructed to describe the connections between between these virtual elements. Due to the unique geometric shape and mass distribution of each virtual element, a non-constant spring parameter is used. This spring parameter has incorporated in it the distance and mass distribution information of the virtual particles. Finally, an eigenvalue decomposition of either Laplacian matrix or Hessian matrix is employed. B-factors and collective modes are then evaluated from eigenvalues and eigenvectors. To facilitate a comparison between my predictions and experimental results, a linear interpolation with the nearest neighbour scheme is used.

The paper is organized as following. Section \ref{sec:Methodology} is devoted for the multiscale virtual particle based elastic network model, including its basic setting, algorithm design and validation. After a brief introduction of Gaussian network model and anisotropic network model in Section \ref{sec:ENM}, I discuss the multiscale representation of biomolecules by our multiscale rigidity function in Section \ref{sec:multiscale_bio}. In Section \ref{sec:MVP}, a detailed discussion of multiscale virtual particle is presented. Multiscale virtual particle model is then combined with Gaussian network model and anisotropic network model. The discussion of MVP-GNM and MVP-ANM is in Section \ref{sec:MVP-GNM} and Section \ref{sec:MVP-ANM}, respectively. In Section \ref{sec:application}, I discuss the collective motion prediction of two virus cases. The paper ends with an conclusion.

\section{Methodology}\label{sec:Methodology}
One of the major features of biological sciences in the 21st century is its transition from an empirical, qualitative and phenomenological discipline to a comprehensive, quantitative and predictive one. Tremendous amount of biomolecular structure data are available in Protein Data Bank and Electron Microscopy Data Bank. The analysis of the dynamics of these structures is essential to the understanding of the function and mechanism of these biomolecules.
Traditional approaches, including molecular dynamics, normal model analysis, elastic network models, etc, have been widely used in biomolecular motion analysis and yielded fruitful results. However, with the availability of more and more large-sized macroproteins and biomolecular complexes, the computational limitations of these approaches begin to reveal. Moreover, the rapid advance in cryo-electron microscopy has enabled researchers to capture structures of extremely large-sized biomolecular complexes and assemblies in their native environment. However, traditional approaches, which can only be applied to structures with atomic details, fall short in the dynamic analysis of cryo-EM data.

To study the dynamic motions of large-sized biomolecular structures, various CG approaches are proposed. As stated in the introduction, these models usually simplify the biomolecular structure graphs/networks and design new potential functions based on them. In this section, a multiscale virtual particle model based elastic network model is proposed to analyze the collective motion of large-sized biomolecules. The MVP-ENM is based on elastic network model and multiscale virtual particle model. It is constructed on the molecular density function, therefore it can be directly applied to Cryo-EM data analysis. And a multiscale representation is used in MVP-ENM. This representation helps to discretize structure data into virtual particles. Details of the method is presented in the following.

\subsection{Elastic network models}\label{sec:ENM}
Due to its great simplicity and efficiency, elastic network model has been widely used in biomolecular flexibility and normal mode analysis \cite{Tirion:1996,Atilgan:2001,Hinsen:1998}. In this model, the potential function used in molecular dynamics has been dramatically simplified. This results in a huge reduction of computational cost. Moreover, a CG representation, i.e., using only $C_{\alpha}$ atoms to generate protein network, is commonly used and has further simplified the model. Gaussian network model \cite{Bahar:1997,Bahar:1998,QCui:2010,LiGH:2002,Yang:2006} and anisotropic network model \cite{Atilgan:2001,Hinsen:1998} are two commonly used ENMs. Generally speaking, GNM is more suitable for biomolecular flexibility analysis. ANM, which generates eigenvector with anisotropic information, can be used in collective motion analysis. Since MVP-ENM is based on elastic network model, I will give a very brief introduction of GNM and ANM in the following sections.

\subsubsection{Gaussian network model (GNM)} \label{sec:GNM}
Gaussian network model is proposed for the study of isotropic fluctuations of biomolecules. For a molecule composed of $N$ residues with $C_{\alpha}$ coordinates ${\bf r}_1, {\bf r}_2,\cdots, {\bf r}_N$, it can deviate from the equilibrium state to a new configuration with $C_{\alpha}$ coordinates represented as ${\bf r}^d_1, {\bf r}^d_2,\cdots, {\bf r}^d_N$. For $C_{\alpha}$ atom $i$ and $j$, at equilibrium state, their distance vector can be represented as ${\bf r}_{ij}={\bf r}_{i}-{\bf r}_{j}$ and distance value is $r_{ij}=|{\bf r}_{ij}|$. At the non-equilibrium state, their vector distance is ${\bf r}^d_{ij}={\bf r}^d_{i}-{\bf r}^d_{j}$ and distance value is $r_{ij}^d=|{\bf r}^d_{ij}|$. With these notations, the Gaussian potential function can be expressed,
\begin{eqnarray}\label{eq:anm_v}
V^{\rm {GNM}}=\gamma \sum_{i,j}^{N} ({ \bf r}^d_{ij}-{\bf r}_{ij})\cdot ({ \bf r}^d_{ij}-{\bf r}_{ij}) f( r_{ij})=\frac{\gamma}{2} \Delta {\bf r} {\bf }^T { L} \Delta {\bf r}.
\end{eqnarray}
Here vector $\Delta {\bf r}$ has $N$ components, and can be represented as $\Delta {\bf r} =( \Delta { r}_{1}, \Delta { r}_{2},...,\Delta { r}_{N} )^T$ with $\Delta { r}_{i}=|{\bf r}^d_{i}-{\bf r}_{i}|, i=1, 2,..., N$. And $T$ denotes the transpose. Heaviside function $f( r_{ij})$ usually involves an interaction cut-off distance $r_c$, and can be expressed as,
\begin{eqnarray}\label{eq:heaviside}
f(r_{ij})=\begin{cases} \begin{array}{ll}
            1 & r_{ij} \leq r_c\\	
            0 & r_{ij} > r_c
	      \end{array}.
\end{cases}
\end{eqnarray}
Laplacian matrix $L$ is a $N$ by $N$ matrix, that can be expressed as,
\begin{eqnarray}\label{eq:couple_matrix25}
L_{ij}=\begin{cases} \begin{array}{ll}
            -1 & i \neq j~{\rm and} ~ r_{ij} \leq r_c\\
			0  & i \neq j~{\rm and} ~ r_{ij} >  r_c\\		
            -\sum_{i \neq j}^N L_{ij} &i=j
	      \end{array}.
\end{cases}
\end{eqnarray}
The equilibrium correlation between fluctuation is,
\begin{eqnarray}\label{eq:gnm_bf1}
<\Delta {\bf r}_{i} \cdot \Delta {\bf r}_{j}>=\frac{3k_BT}{\gamma} (L^{-1})_{ij}, ~\forall i=1,2,\cdots, N,
\end{eqnarray}
Here $L^{-1}$ is the Moore-Penrose pseudo-inverse of Laplacian matrix $L$. More specifically,
$$\left(L^{-1} \right)_{ii}=\sum_{k=2}^N  \lambda_k^{-1}\left[{\bf v}_k {\bf v}_k^T \right]_{ii},$$
where $T$ denotes the transpose and $\lambda_k$ and ${\bf v}_k$ are the $k$-th eigenvalue and eigenvector of $\Gamma$, respectively. The summation omits the first eignmode, as its eigenvalue is zero.

Further, GNM prediction of the $i$-th B-factor of the biomolecule can be expressed as \cite{Bahar:1997,Bahar:1998,JKPark:2013}
\begin{eqnarray}\label{eq:gnm_bf2}
B_i^{\rm GNM}=\frac{8 \pi^2}{3} <\Delta {\bf r}_{i} \cdot \Delta {\bf r}_{i}>, ~\forall i=1,2,\cdots, N.
\end{eqnarray}

The experimental B-factor values can be found in the PDB data. It is an indication of the relative thermal fluctuations of atoms. Atoms with small B-factors belong to a part of the structure that is very rigid. Atoms with large B-factors generally belong to part of the structure that is very flexible. GNM has proved to be a highly efficient and accurate method for B-factor prediction. It has been demonstrated by Yang et al. \cite{LWYang:2008} that the GNM is about one order more efficient than most other flexibility approaches. It is worth mentioning that the graph or network in GNM is constructed by using a cutoff distance $r_c$. Usually, the cutoff distance value is chosen as $8$\AA.

\subsubsection{Anisotropic network model (ANM)}\label{sec:ANM}
As state above, GNM uses only distance information with no consideration of anisotropic properties. To restore the anisotropic fluctuations, ANM has been proposed. Similar to GNM, a CG representation is usually employed with each residue represented by its $C_{\alpha}$ atom.  To faciliate the explanation, the same coordinate notations used in the previous section is used. Further, we denote ${\bf r}_i=(x_i, y_i, z_i)$, ${\bf r}_i^d=(x_i^d, y_i^d, z_i^d)$ for $i=1, 2, ..., N$, and
$\Delta {\bf R}= \{ \Delta x_1, \Delta y_1, \Delta z_1,...,\Delta x_N, \Delta y_N, \Delta z_N \}$ with $\Delta x_i=x_i^d-x_i$, $\Delta y_i=y_i^d-y_i$ and $\Delta z_i=z_i^d-z_i$ for $i=1, 2, ..., N$. The ANM potential function can be expressed,
\begin{eqnarray}\label{eq:anm_v}
V^{\rm {ANM}}=\gamma \sum_{i,j}^{N} ({ r}^d_{ij}-{ r}_{ij})^2 f( r_{ij})=\frac{\gamma}{2} \Delta {\bf R}^T { H} \Delta {\bf R}.
\end{eqnarray}
Here $r_{ij}=|{\bf r}_{ij}|$ and $r_{ij}^d=|{\bf r}_{ij}^d|$ are distances between between $i$-th and $j$-th atoms in their equilibrium and non-equilibrium states. Heaviside function $f( r_{ij})$ is the same as in Eq. (\ref{eq:heaviside}), except that a large cutoff distance is used. Usually, it is chosen as $r_c=12$ \AA.~
And Hessian matrix $H$ is $3N$ by $3N$ matrix, which composes many local $3$ by $3$ off-diagonal matrix $H_{ij}$ as following,
\begin{eqnarray}\label{eq:multi-kirchoff1}
 H_{ij} = -\frac{1}{r_{ij}^2}\left[ \begin{array}{ccc}
	        (x_j-x_i)(x_j-x_i) & (x_j-x_i)(y_j-y_i) & (x_j-x_i)(z_j-z_i)\\
             (y_j-y_i)(x_j-x_i) & (y_j-y_i)(y_j-y_i) & (y_j-y_i)(z_j-z_i)\\
             (z_j-z_i)(x_j-x_i) & (z_j-z_i)(y_j-y_i) & (z_j-z_i)(z_j-z_i)
	      \end{array}\right]  ~ i,j=1,2,\cdots, N, i \neq j ~{\rm and} ~ r_{ij}\leq r_c.
 \end{eqnarray}
Here the coordinates for $i$-th and $j$-th atoms are ${\bf r}_{i}=(x_i,y_i,z_i)$ and ${\bf r}_{j}=(x_j,y_j,z_j)$. The $r_{ij}=||{\bf r}_{i}-{\bf r}_{j}||$ is the distance between them.
Same as in the GNM, the diagonal part is the negative summation of the off diagonal elements:
\begin{eqnarray}\label{eq:multi-kirchoff1_diagonal}
 H_{ii} = -\sum_{i\neq j} H_{ij}, ~\forall i=1,2,\cdots, N.
 \end{eqnarray}
Compared with GNM, which uses a Laplacian matrix with size $N$ by $N$, the Hessian matrix constructed in ANM is $3N$ by $3N$. After eigenvalue decomposition, the resulting eigenvector has $3N$ components, i.e., each atom contributes three components in an eigenvector. These components, known as the eigenmode, give the directions for atoms to move along with in ${\mathbb R}^3$.


The equilibrium correlation between fluctuation is,
\begin{eqnarray}\label{eq:anm_bf1}
<\Delta {\bf R}_{i} \cdot \Delta {\bf R}_{j}>=\frac{3k_BT}{\gamma} (H^{-1})_{ij}, ~\forall i,j=1, 2, \cdots, 3N.
\end{eqnarray}
Again $H^{-1} $ is the Moore-Penrose pseudo-inverse of Hessian matrix $H$. More specifically, $$\left(H^{-1} \right)_{ii}=\sum_{k=7}^{3N}  \lambda_k^{-1}\left[{\bf v}_k {\bf v}_k^T \right]_{ii},$$
where $T$ denotes the transpose and $\lambda_k$ and ${\bf v}_k$ are the $k$-th eigenvalue and eigenvector of $H$, respectively. The summation omits the first six eignmodes, whose eigenvalues are zero.

Further, ANM prediction of the $i$-th B-factor of the biomolecule can be expressed as \cite{Bahar:1997,Bahar:1998,JKPark:2013}
\begin{eqnarray}\label{eq:anm_bf2}
B_i^{\rm ANM}=\frac{8 \pi^2}{3} \sum^{3i}_{j=3i-2} < \Delta {\bf R}_{j} \cdot \Delta {\bf R}_{j}>, ~\forall i=1, 2,\cdots, N.
\end{eqnarray}

ANM has proved to be a powerful tool for normal mode analysis. Due to its great efficiency, it has been successfully applied to macroproteins and protein complexes, such as, Ribosomes \cite{Tama:2003,YWang:2004}, Virus capsid\cite{Rader:2005,Tama:2005}, chaperonin GroEL\cite{WZheng:2007}, etc. Various softwares and online solvers have been built up\cite{Skjaven:2009}. A more detailed discussion can be found in these review papers\cite{LWYang:2008,QCui:2010}.

\subsection{Multiscale representation of biomolecules}\label{sec:multiscale_bio}
Biomolecular data are usually highly complicated and essentially multiscale. Our previous works on multiscale models have indicated that a better representation of the multiscale biomolecular structure can improve the accuracy of the models in flexility analysis\cite{Opron:2015communication,Xia:2015multiscale,Nguyen:2016generalized}. And a multiscale rigidity function has been proposed. The essential idea for this function is to match the scale of interest with appropriate resolution. Simply speaking, a resolution parameter is introduced into my multiscale rigidity function, and by turning this parameter, a series of representations in various scales can be generated. Suitable representations can be chosen depending on the scale of interest. Mathematically, this is achieved  by converting a discrete point cloud data into a series of continuous density functions. 

To be more specific, for a data set with a total $N$ entries, which can be atoms, residues, domains and protein monomers, if one assumes their generalized coordinates are ${\bf r}_1, {\bf r}_2,\cdots, {\bf r}_N$, a multiscale rigidity function of the data can be expressed as,
\begin{eqnarray}\label{eq:rigidity_function}
\mu({\bf r},\eta)=\sum_{j}^N w_j\Phi(\parallel {\bf r}- {\bf r}_{j}\parallel;\eta),
\end{eqnarray}
where $w_j$ is $j$-th weight parameter. For example, it can be chosen as the atomic number. The parameter $\eta$ is the resolution or scale parameter. The function $\Phi(\parallel {\bf r}- {\bf r}_{j}\parallel;\eta)$ is a kernel function. Commonly used kernel functions include  generalized exponential functions,
\begin{eqnarray}\label{eq:couple_matrix1}
\Phi(\parallel {\bf r}- {\bf r}_{j}\parallel ;\eta, \kappa) =    e^{-\left( \parallel {\bf r}- {\bf r}_{j}\parallel /\eta \right)^\kappa},    \quad \kappa >0
\end{eqnarray}
or generalized Lorentz functions,
\begin{eqnarray}\label{eq:couple_matrix2}
 \Phi(\parallel {\bf r}- {\bf r}_{j}\parallel;\eta, \upsilon) =  \frac{1}{1+ \left( \parallel {\bf r}- {\bf r}_{j}\parallel /\eta \right)^{\upsilon}},  \quad  \upsilon >0.
 \end{eqnarray}
It can be noticed that the larger $\eta$ value is, the lower its resolution is.

A multiscale geometric model can be naturally derived from this multiscale rigidity functions. By using different resolution values, these models can capture structure properties in different scales. To further facilitate the comparison between these different scales, a normalized density function is proposed as following,
\begin{eqnarray}\label{eq:scaled_rigidity_function}
\mu^s({\bf r};\eta)=\frac{\mu({\bf r};\eta)-\mu_{\min}}{\mu_{\max}-\mu_{\min}}.
\end{eqnarray}
Here $\mu_{\max}$ and $\mu_{\min}$ are the maximum and minimum of the density function. The values of the normalized multiscale function is within the range [0, 1]. It is worth mentioning that my MVP-ENM is based on density or mass distributions. Therefore, for structures derived from X-ray and NMR, their density functions and the normalized multiscal rigidity function will be generated before firstly.

After this brief introduction of Gaussian network model, anisotropic network model and our multiscale rigidity function, now I am ready to introduce the MVP-ENM.

\subsection{Multiscale virtual particle based elastic network model}\label{sec:MVP-ENM}
As stated in the introduction, a pronounced weakness of traditional NMA, ENM, and ANM is that they can not by used in Cryo-EM data analysis\cite{Ming:2002describe}. State differently, they are designed based on atomic structures and can not be directly applied to continuous density functions. To overcome this problem, quantized elastic deformation model is proposed. In this method, vector quantization algorithm \cite{Gray:1984vector} is employed to decompose the electron density map of a biological molecule into a set of finite Voronoi cells. It is then combined with ANM to explore the dynamic of the cryo-EM data \cite{Tama:2002exploring,Ming:2002describe}. This model has been proved to be consistent with ANM results and be able to predict the experimental B-factors with a high accuracy.

Motivated by the success of QEDM, I propose a more generalized elastic normal mode method for the collective motion analysis of biomolecular density data. The essence of my MVP-ENM is to discretize a molecular density function into multiscale virtual particles and then ``connect" them with non-uniformed springs. Originally, virtual particles are introduced for the discretization of molecular density data \cite{Xia:2017multiscale}. Since there is no explicit atom coordinates, virtual particles work as building blocks of the molecular. A harmonic potential is assumed between any two virtual particles. The potential value depends on not only the relative positions but also the density distributions. More specifically, a non-uniformed spring parameter, which incorporates both geometric and physical properties of virtual particles, is used in the harmonic potential function. Further, a multiscale representation is considered. In this way, MVP-ENM can work as a multiscale coarse-grained framework for normal mode analysis. A detailed description is presented below.


\subsubsection{Multiscale virtual particle model}\label{sec:MVP}

\begin{figure}
\begin{center}
\begin{tabular}{c}
\includegraphics[width=0.8\textwidth]{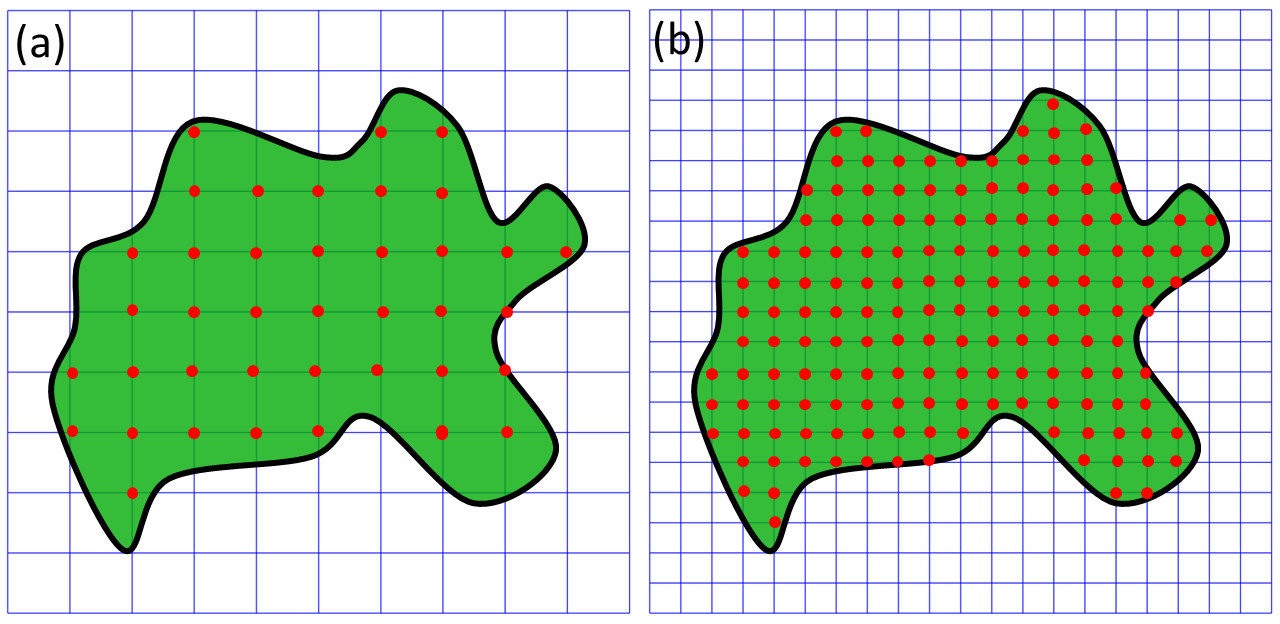}
\end{tabular}
\end{center}
\caption{The illustration of multiscale virtual particle models. A two dimensional slice from the three dimensional data is demonstrated. The protein domain and its boundary are marked by green color and black color, respectively. The Cartesian grid is used to do the discretization. Two different virtual particle models are presented. The centers of virtual particles are represented by red dots. It can be seen that two models capture the basic geometry of the biomolecule in different resolutions.
}
\label{fig:virtual_particle}
\end{figure}

The multiscale virtual particle model includes two essential components. The first one is to generate a virtual particle model in suitable scales, based on the biomolecular density data. The suitable scales mean that they are able to balance between the accuracy of the structure representation and the computational complexity. The other component is the special spring parameter that reflects both distance and density relation between two virtual particles. Essentially, a soft kernel is used to replace the cutoff distance in traditional ENM.

\paragraph{Generation of multiscale virtual particles}
In our MVP model, the computational domain of a density distribution function is defined as the inner region/regions enclosed by a certain boundary. This boundary specifies the shape or surface of the biomolecule. It is usually generated by using a certain isovalue or a level set value. The computational domain can be discretized in various ways. Tetrahedron mesh, Cartesian grid, Voronoi diagram, and other domain decomposition methods from numerical methods can all be used. Virtual particles can be very irregular in terms of their sizes and shapes. Density distributions can also vary from particle to particle.

More importantly, the discretization can be done in various scales and the resulting virtual particle model can have different accuracies in terms of structure representation. Similar to the domain decomposition in numerical methods, to select a coarser grid or a more refined grid depends on the affordable computational power. Figure \ref{fig:virtual_particle} demonstrates virtual particle models in two different scales. The Cartesian grid is used in our model. I only depict a two dimensional slice from the three dimensional data. The protein and its boundary are marked by green color and black color, respectively. Virtual particles are represented by red dots. It can be seen that both models capture the basic geometry of the biomolecule. Further, since the density distribution naturally incorporates in it the biomolecular geometric details, a systematical sampling within the biomolecular domain can preserve this information very well.

\paragraph{Connection between multiscale virtual particles}
The most critical piece of the MVP model is to find suitable ``connections" between virtual particles. Since the virtual particles vary greatly in their shapes, sizes and mass-distributions, an identical spring constant as in the traditional elastic network models will no longer be suitable. Therefore, we proposed a new spring parameter that is determined by particle properties\cite{Xia:2017multiscale}. For example, if a biomolecule has a normalized density distribution $\mu^s({\bf r})$, the spring parameter for its two virtual particles centered at ${\bf r}_I$ and ${\bf r}_J$ and enclosed by the volume elements of $\Omega_I$ and $\Omega_J$, can be expressed as following:
\begin{eqnarray}\label{rigidity_potential8}
 \gamma({\bf r}_I,{\bf r}_J,\Omega_I,\Omega_J,\mu^s({\bf r}),\eta^{\rm MVP})= \gamma_1(\Omega_I,\Omega_J,\mu^s({\bf r}))\cdot  \gamma_2({\bf r}_I,{\bf r}_J,\eta^{\rm MVP})
\end{eqnarray}
Here $\gamma_1(\Omega_I,\Omega_J,\mu^s({\bf r}))$ is the mass or density contribution to the spring parameter. And $\gamma_2({\bf r}_I,{\bf r}_J,\eta^{\rm MVP})$ is the distance influence on spring parameter. The resolution parameter $\eta^{\rm MVP}$ is related to the scale of the MVP model.

For the density contribution part, since the spring parameter between two virtual particles is positively related to their total density, virtual particles with large total density will have a large spring constant. With this consideration, $\gamma_1(\Omega_I,\Omega_J,\mu^s({\bf r}))$ can be modeled as,
\begin{eqnarray}\label{Eq:S1_F1}
\gamma_1(\Omega_I,\Omega_J,\mu^s({\bf r}))=\left(1+a\int_{\Omega_I}\mu^s({\bf r}) d{\bf r}\right)\left(1+a\int_{\Omega_J}\mu^s({\bf r}) d{\bf r}\right),
\end{eqnarray}
or
\begin{eqnarray}\label{Eq:S1_F2}
\gamma_1(\Omega_I,\Omega_J,\mu^s({\bf r}))=a \left(\int_{\Omega_I}\mu^s({\bf r}) d{\bf r}+ \int_{\Omega_J}\mu^s({\bf r}) d{\bf r} \right).
\end{eqnarray}
The parameter $a$ is a weight value.

For the distance contribution part, a distance related function is considered in my model. Traditional elastic network models usually employ a cutoff distance. Atoms within this cutoff distance are connected by a uniformed spring value. In our virtual particle model, the cutoff distance is replaced by a soft kernel. Essentially, distance related parameter $\gamma_2({\bf r}_I,{\bf r}_J,\eta^{\rm MVP})$ can be chosen from generalized Gaussian kernels or Lorentz kernels as in Eq.(\ref{eq:couple_matrix1}) and Eq.(\ref{eq:couple_matrix2}). For example,
\begin{eqnarray}\label{Eq:S2}
\gamma_2({\bf r}_I,{\bf r}_J,\eta^{\rm MVP}) =  e^{-\left( \parallel {\bf r}_I- {\bf r}_{J}\parallel /\eta^{\rm MVP} \right)^\kappa},  \quad \kappa >0.
\end{eqnarray}
The value of the scale parameter $\eta^{\rm MVP}$ depends on the multiscale virtual particle model. Normally, its value should be around the size of the virtual particle.

With the basic setting of multiscale virtual particle model, I am ready to introduce the MVP-ENM, which includes two different models, i.e., a multiscale virtual particle based Gaussian network model and a multiscale virtual particle based anisotropic network model.

\subsubsection{Multiscale virtual particle based Gaussian network model (MVP-GNM)}\label{sec:MVP-GNM}

\begin{figure}
\begin{center}
\begin{tabular}{c}
\includegraphics[width=0.8\textwidth]{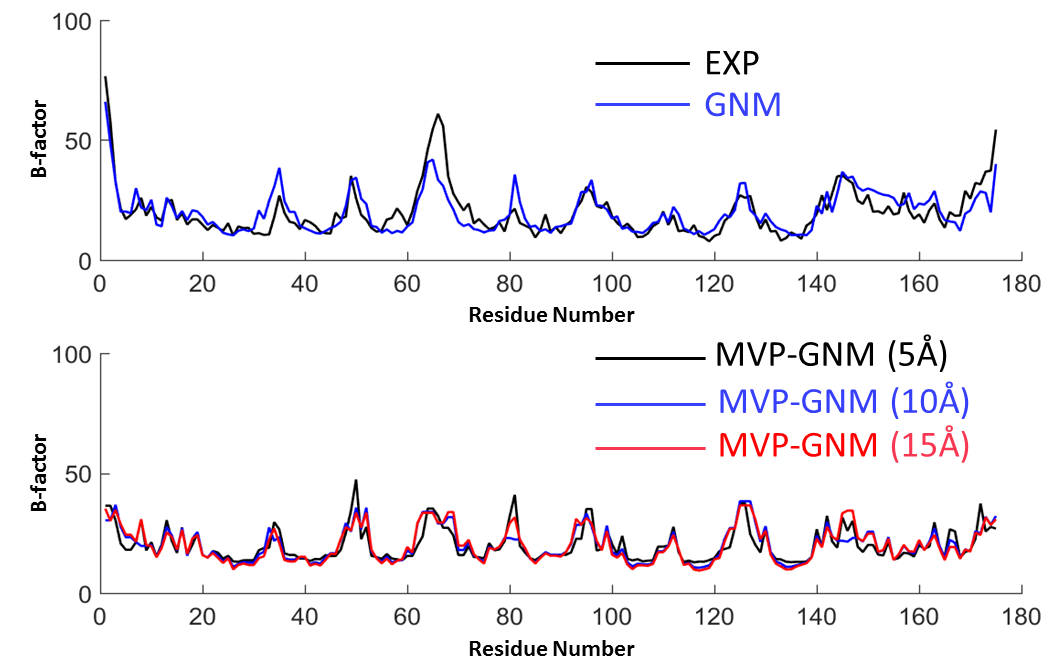}
\end{tabular}
\end{center}
\caption{ Comparison of B-factor prediction with GNM and MVP-GNM for protein 1AQB. Three different scale values are used $\sigma= 5$\AA, $10$\AA~and $15$\AA. The PCC for GNM is 0.822, and PCCs for VP-GNM with $\sigma= 5$\AA, $10$\AA~and $15$\AA~ are 0.666, 0.657 and 0.699, respectively.
}
\label{fig:1AQB_bf_gnm}
\end{figure}

\begin{figure}
\begin{center}
\begin{tabular}{c}
\includegraphics[width=0.8\textwidth]{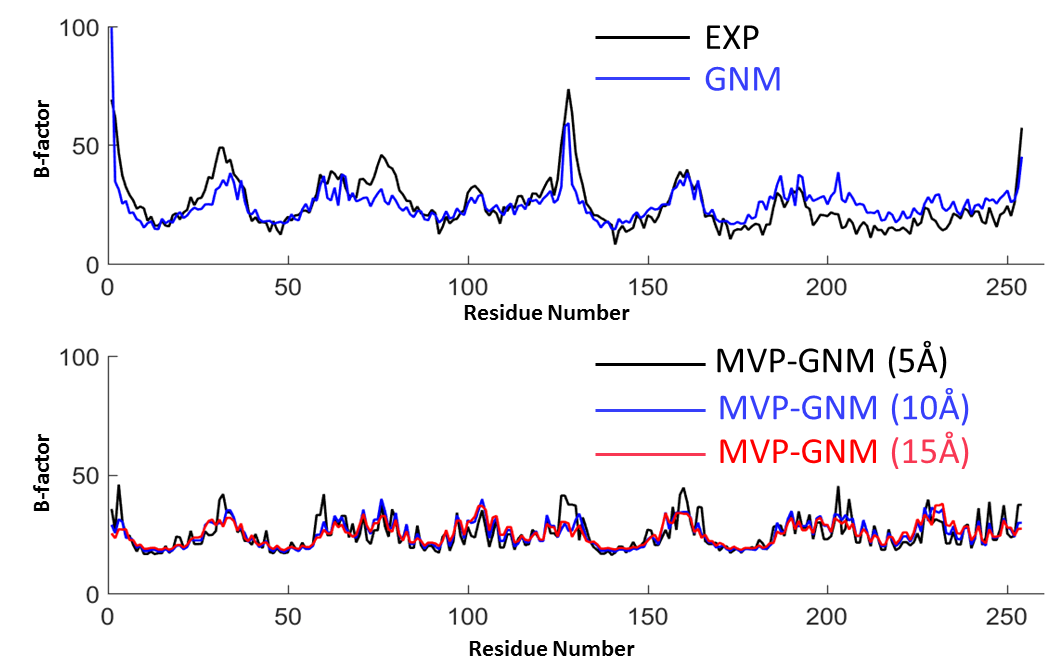}
\end{tabular}
\end{center}
\caption{ Comparison of B-factor prediction with GNM and MVP-GNM for protein 2CCY. Three different scale values are used $\sigma= 5$\AA, $10$\AA~and $15$\AA. The PCC for GNM is 0.739, and PCCs for VP-GNM with $\sigma= 5$\AA, $10$\AA~and $15$\AA~ are 0.623, 0.507 and 0.439, respectively.
}
\label{fig:2CCY_bf_gnm}
\end{figure}

In our multiscale virtual particle based Gaussian network model, a new potential function can be expressed as following,
\begin{eqnarray}\label{eq:gnm_v}
V^{\rm {MVP-GNM}}=\frac{1}{2} \Delta {\bf r}^T L^{\rm {MVP-GNM}} \Delta {\bf r}.
\end{eqnarray}
Here $L^{\rm {MVP-GNM}}$ is a Laplacian matrix, which has incorporated in it the spring parameter information. More specifically, it can be expressed as,
\begin{eqnarray}\label{eq:couple_matrix25}
L_{ij}^{\rm {MVP-GNM}}=\begin{cases} \begin{array}{ll}
            -\gamma({\bf r}_I,{\bf r}_J,\Omega_I,\Omega_J,\mu^s(r)) & i \neq j\\		
            -\sum_{i \neq j}^N L_{ij} &i=j
	      \end{array}.
\end{cases}
\end{eqnarray}
The formula for the equilibrium correlation between fluctuation and B-factor prediction remain the same as in Eq.(\ref{eq:gnm_bf1}) and Eq.(\ref{eq:gnm_bf2}), respectively. However, their values are calculated not on atoms, but on virtual particles. To facilitate a comparison between our MVP-GNM and traditional GNM, the B-factor value of an atom is interpolated from the calculated B-factors on the virtual particles with the nearest neighbour formula.

To validate our MVP-GNM model, I choose three proteins from the protein data bank. Their IDs are 2CCY, 1AQB and 2ABH. The molecular density data is generated by using the generalized Gaussian model in Eq. (\ref{eq:couple_matrix1}) with $\kappa=2$. Three different $\eta$ values, i.e., $\eta=5$\AA, $10$\AA~ and $15$\AA~, are employed in our test cases to simulate molecular density data in different scales. In our discretization schemes, the Cartesian grid with a grid spacing $4.0$\AA~ is used. The protein domain is chosen as the regions with normalized density value larger than or equal to $0.4$, i.e., $\mu^s({\bf r};\eta) \geq 0.4$. In the GNM model, the cutoff distance is chosen as the commonly used value 7\AA.

For spring parameter $\gamma_1(\Omega_I,\Omega_J,\mu^s({\bf r}))$, Eq.(\ref{Eq:S1_F1}) is used with weight value $a=1$. For $\gamma_2({\bf r}_I,{\bf r}_J,\eta^{\rm MVP})$, the generalized Gaussian kernel as in Eq. (\ref{Eq:S2}) is used. The parameter $\kappa$ is chosen as $2$ and scale parameter $\eta^{\rm MVP}$ is chosen as the same value as scale parameter $\eta$.

\begin{figure}
\begin{center}
\begin{tabular}{c}
\includegraphics[width=0.8\textwidth]{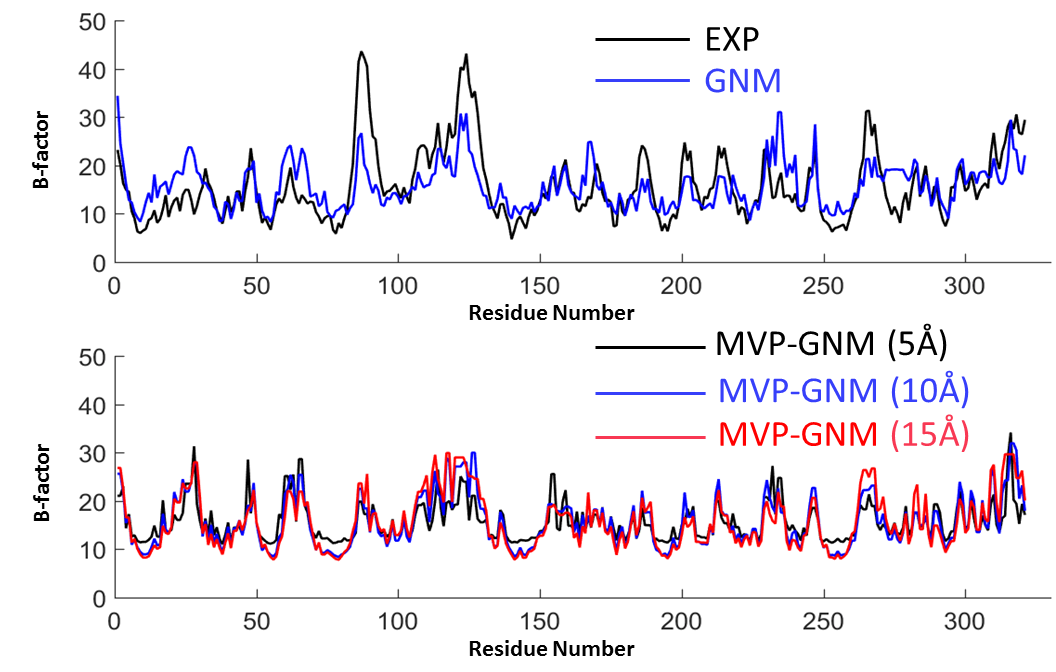}
\end{tabular}
\end{center}
\caption{ Comparison of B-factor prediction with GNM and MVP-GNM for protein 2ABH. Three different scale values are used $\sigma= 5$\AA, $10$\AA~and $15$\AA. The PCC for GNM is 0.647, and PCCs for VP-GNM with $\sigma= 5$\AA, $10$\AA~and $15$\AA~ are 0.550, 0.731 and 0.775, respectively.
}
\label{fig:2ABH_bf_gnm}
\end{figure}

The B-factor prediction results of our MVP-GNM and GNM on three proteins are illustrated in Figures \ref{fig:1AQB_bf_gnm} to \ref{fig:2ABH_bf_gnm}. For protein 1AQB, the Pearson correlation coefficients (PCCs) between experimental B-factors and the predicted B-factors by GNM is 0.822. And PCCs between experimental B-factors and the predicted B-factors by MVP-GNM with $\sigma= 5$\AA, $10$\AA~and $15$\AA~ are 0.623, 0.507 and 0.439, respectively. Even though our MVP-GNM does not provide a better results than GNM, it captures the basic B-factor profile very well. By the comparison of the experimental results and our predictions, one can find that the mismatch between them comes largely from the regions with extremely large B-factor values.

For protein 2CCY, the PCCs between experimental B-factors and the predicted B-factors by GNM is 0.739. And PCCs between experimental B-factors and the predicted B-factors by MVP-GNM with $\sigma= 5$\AA, $10$\AA~and $15$\AA~ are  0.623, 0.507 and 0.439, respectively. Again, our prediction is able to preserve the basic pattern of the original B-factor profile, and the mismatch comes from regions with B-factor value peaks.

For protein 2ABH, the PCCs between experimental B-factors and the predicted B-factors by GNM is 0.647. And PCCs between experimental B-factors and the predicted B-factors by MVP-GNM with $\sigma= 5$\AA, $10$\AA~and $15$\AA~ are  0.550, 0.731 and 0.775, respectively. Again, our prediction is able to preserve the basic pattern of the original B-factor profile. The results from our MVP-GNM at $\sigma=10$\AA~and $15$\AA~ are better than the results from the GNM. Moreover, the mismatch still comes from regions with B-factor value peaks.

From the above examples, it can be seen that my MVP-GNM is essentially a generalization of GNM from the atomic coordinate representation to an ``arbitrary" discrete point set representation. Actually, if virtual particles are specially designed to match atom centers, MVP-GNM goes back to the GNM. Except that this GNM uses a soft kernel instead of a cutoff distance. It should be noticed that MVP-GNM delivers inferior results than GNM largely due to two reasons. The first reason is that the grid spacing in MVP-GNM is relatively large. A grid spacing of 4 \AA~ is not good enough to capture all the necessary structure details, as the distance between two $C_{\alpha}$ is 3.8 \AA. The other reason is that the resolution parameter is also relative large. Actually, the smallest resolution parameter is $5$\AA.~Again relevant structure details tend to be smoothed out in this resolution. However, even with this coarse grid spacing and coarse structure representation, MVP-GNM is still able to predict the intrinsic fluctuations nicely. Since the mismatches unanimously come from the large fluctuation regions, this means that the MVP-GNM captures low frequency mode information very accurately. The results are highly consistent with previous findings that flexibilities are highly related to local packing density\cite{Halle:2002}. More importantly, both flexibility and collective modes are largely determined by biomolecular shapes and density distributions\cite{Ming:2002describe,chacon2003mega}.

\subsubsection{Multiscale virtual particle based anisotropic network model (MVP-ANM) }\label{sec:MVP-ANM}

In our multiscale virtual particle based anisotropic network model, a new potential function can be expressed as following,
\begin{eqnarray}\label{eq:anm_v}
V^{\rm {MVP-ANM}}=\frac{1}{2} \Delta {\bf R}^T {H}^{\rm {MVP-ANM}} \Delta {\bf R}
\end{eqnarray}
Here ${H}^{\rm {MVP-ANM}}$ is the new Hessian matrix, which has incorporated in it the spring parameter information. Hessian matrix ${H}^{\rm {MVP-ANM}}$ is still $3N$ by $3N$, which composes many local $3$ by $3$ off-diagonal matrix as following,
\begin{eqnarray}\label{eq:multi-kirchoff1}
 H_{IJ}^{\rm {MVP-ANM}} = -\frac{\gamma_{IJ}}{r_{ij}^2}\left[ \begin{array}{ccc}
	        (x_J-x_I)(x_J-x_I) & (x_J-x_I)(y_J-y_I) & (x_J-x_I)(z_J-z_I)\\
             (y_J-y_I)(x_J-x_I) & (y_J-y_I)(y_J-y_I) & (y_J-y_I)(z_J-z_I)\\
             (z_J-z_I)(x_J-x_I) & (z_J-z_I)(y_J-y_I) & (z_J-z_I)(z_J-z_I)
	      \end{array}\right]  ~I \neq J.
 \end{eqnarray}
Here $\gamma_{IJ}=\gamma({\bf r}_I,{\bf r}_J,\Omega_I,\Omega_J,\mu^s({\bf r}),\eta^{\rm MVP})$ is the spring parameter between virtual particles at $r_I$ and $r_J$. The diagonal part is the negative summation of the off-diagonal elements:
\begin{eqnarray}\label{eq:multi-kirchoff1_diagonal}
 H_{II}^{\rm {MVP-ANM}} = -\sum_{I\neq J} H_{IJ}^{\rm {MVP-ANM}}, ~\forall i=1,2,\cdots, N.
 \end{eqnarray}
The formula for equilibrium correlation between fluctuation and B-factor prediction remains the same as in Eq.(\ref{eq:anm_bf1}) and Eq.(\ref{eq:anm_bf2}), respectively. Similar to MVP-GNM, their values are calculated not on atoms, but on virtual particles. To facilitate a comparison between my MVP-ANM and traditional ANM, B-factors and eigenmodes are interpolated from those predicted on virtual particles to atoms using the nearest neighbour formula.

\paragraph{B-factor prediction}

\begin{figure}
\begin{center}
\begin{tabular}{c}
\includegraphics[width=0.8\textwidth]{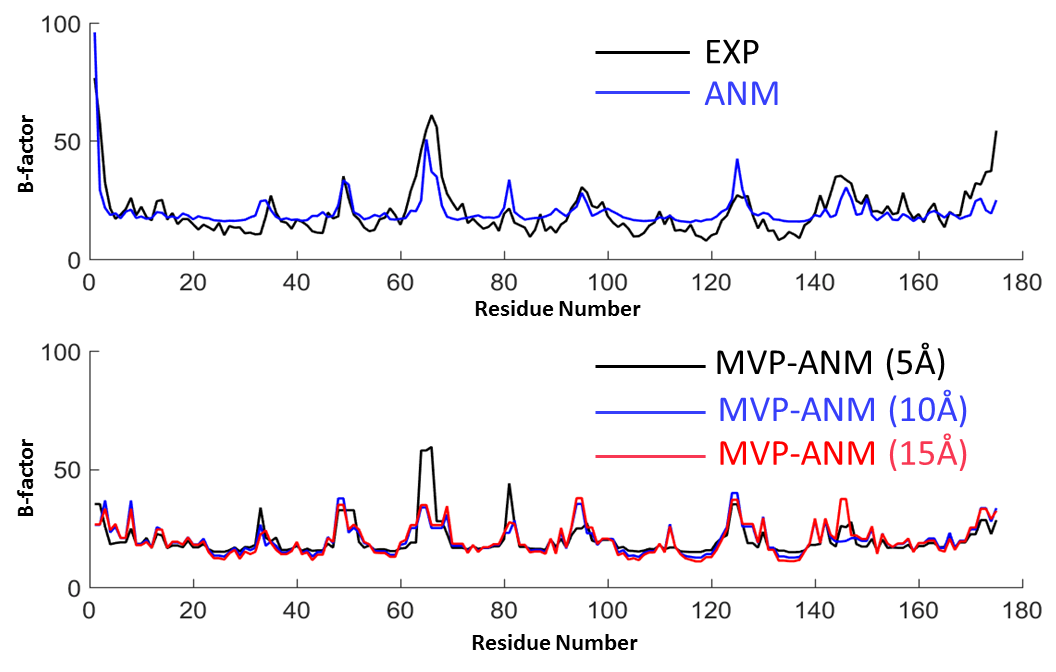}
\end{tabular}
\end{center}
\caption{ Comparison of B-factor prediction with ANM and MVP-ANM for protein 1AQB. Three different scale values are used $\sigma= 5$\AA, $10$\AA~and $15$\AA. The PCC between ANM predictions and experimental results is 0.725, and PCCs for MVP-ANM with $\sigma= 5$\AA, $10$\AA~and $15$\AA~ are 0.696, 0.593 and 0.646, respectively.
}
\label{fig:1AQB_bf_anm}
\end{figure}

\begin{figure}
\begin{center}
\begin{tabular}{c}
\includegraphics[width=0.8\textwidth]{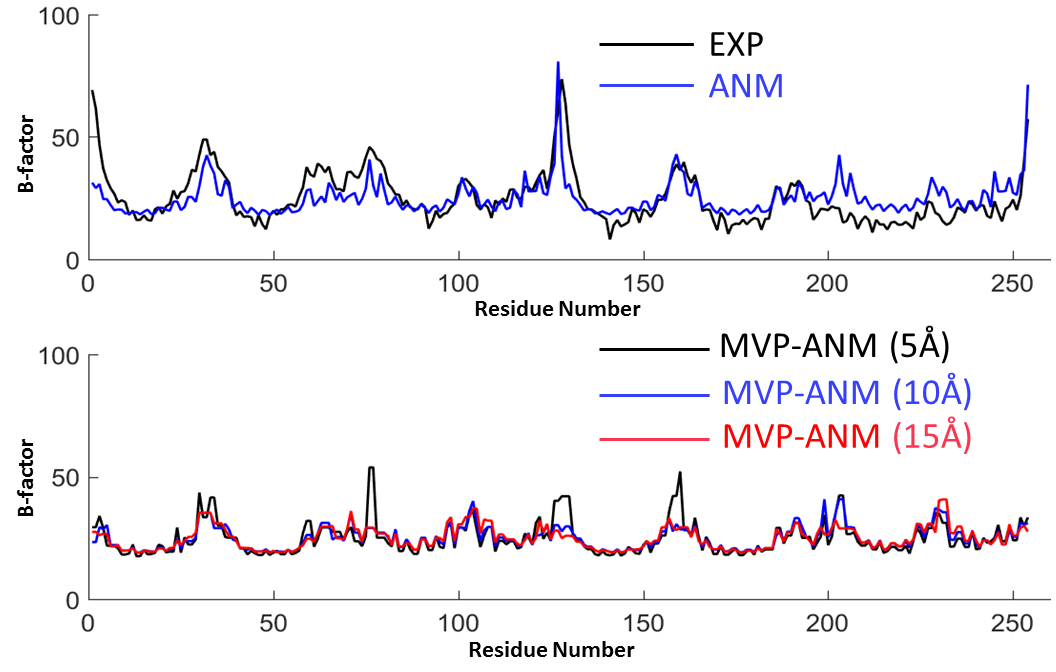}
\end{tabular}
\end{center}
\caption{ Comparison of B-factor prediction with ANM and MVP-ANM for protein 2CCY. Three different scale values are used $\sigma= 5$\AA, $10$\AA~and $15$\AA. The PCC between ANM predictions and experimental results is 0.664 , and PCCs for MVP-ANM with $\sigma= 5$\AA, $10$\AA~and $15$\AA~ are 0.627, 0.450 and 0.435, respectively.
}
\label{fig:2CCY_bf_anm}
\end{figure}

\begin{figure}
\begin{center}
\begin{tabular}{c}
\includegraphics[width=0.8\textwidth]{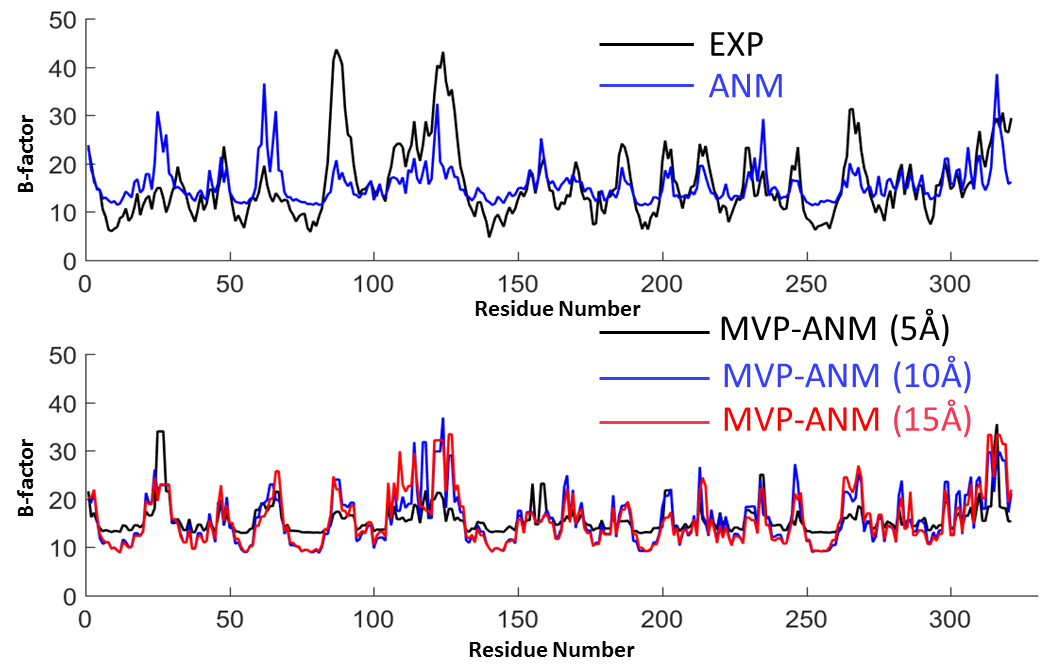}
\end{tabular}
\end{center}
\caption{ Comparison of B-factor prediction with ANM and MVP-ANM for protein 2ABH. Three different scale values are used $\sigma= 5$\AA, $10$\AA~and $15$\AA. The PCC between ANM predictions and experimental results is 0.548, and PCCs for MVP-ANM with $\sigma= 5$\AA, $10$\AA~and $15$\AA~ are 0.442, 0.743 and 0.760, respectively.
}
\label{fig:2ABH_bf_anm}
\end{figure}

To validate MVP-ANM, I compare the B-factor prediction results of the original ANM and my MVP-ANM. The three proteins used in the previous MVP-GNM are still considered. In ANM, a cutoff distance of 12 \AA~ is used. The molecular density data is generated with the same kernel as in the MVP-GNM cases, i.e., the generalized Gaussian model in Eq. (\ref{eq:couple_matrix1}) with $\kappa=2$ and three different $\eta$ values $\eta=5$\AA, $10$\AA~ and $15$\AA. The protein domain is still chosen as the regions with normalized density value larger than or equal to $0.4$. The only difference is that, in my MVP-ANM, the grid spacing is $5.0$\AA. Again, to facilitate a better comparison, the calculated B-factors on the virtual particles are projected onto the atomic centers by using the nearest neighbour interpolation. The B-factor prediction results of our MVP-ANM and ANM on three proteins are illustrated in Figures \ref{fig:1AQB_bf_anm} to \ref{fig:2ABH_bf_anm}.

For protein 1AQB, the PCCs between experimental B-factors and the predicted B-factors by ANM is 0.725. And PCCs between experimental B-factors and the predicted B-factors by MVP-ANM with $\sigma= 5$\AA, $10$\AA~and $15$\AA~ are 0.696, 0.593 and 0.646, respectively. Even though our MVP-ANM does not provide a better results than ANM, it captures the basic B-factor profile very well. By the comparison of the experimental results and our predictions, one can find that the mismatch between them comes largely from the regions with extremely large B-factor values.

For protein 2CCY, the PCCs between experimental B-factors and the predicted B-factors by ANM is 0.664. And PCCs between experimental B-factors and the predicted B-factors by MVP-ANM with $\sigma= 5$\AA, $10$\AA~and $15$\AA~ are  0.627, 0.450 and 0.435, respectively. MVP-ANM is able to preserve the basic pattern of the original B-factor profile. And the mismatch again comes from regions with B-factor value peaks.

For protein 2ABH, the PCCs between experimental B-factors and the predicted B-factors by ANM is 0.548. And PCCs between experimental B-factors and the predicted B-factors by MVP-ANM with $\sigma= 5$\AA, $10$\AA~and $15$\AA~ are  0.442, 0.743 and 0.760, respectively. Again, MVP-ANM is able to preserve the basic pattern of the original B-factor profile. The results from our MVP-ANM at $\sigma=10$\AA~and $15$\AA~ are even better that the results from the ANM. Moreover, the mismatch still comes from regions with B-factor value peaks.

From the above results, it can be seen that my MVP-ANM is able to preserve the global properties of B-factors very well. These global properties are directly related to the general shape of the biomolecular structure and can be determined by low-frequency modes. Even though my MVP-ANM uses a coarse structure representation with a coarse grid, the general structure information is well preserved in the model. So that comparably good results can be obtained.

Further, by the comparison of the results from MVP-GNM and MVP-ANM, it can be seen that MVP-ENMs are essentially a generalization of ENM from the atomic coordinate representation to a density related point cloud representation. This model has demonstrated a great performance by maintaining the global properties even in very coarse representations.

\paragraph{Normal mode prediction}

\begin{figure}
\begin{center}
\begin{tabular}{c}
\includegraphics[width=0.7\textwidth]{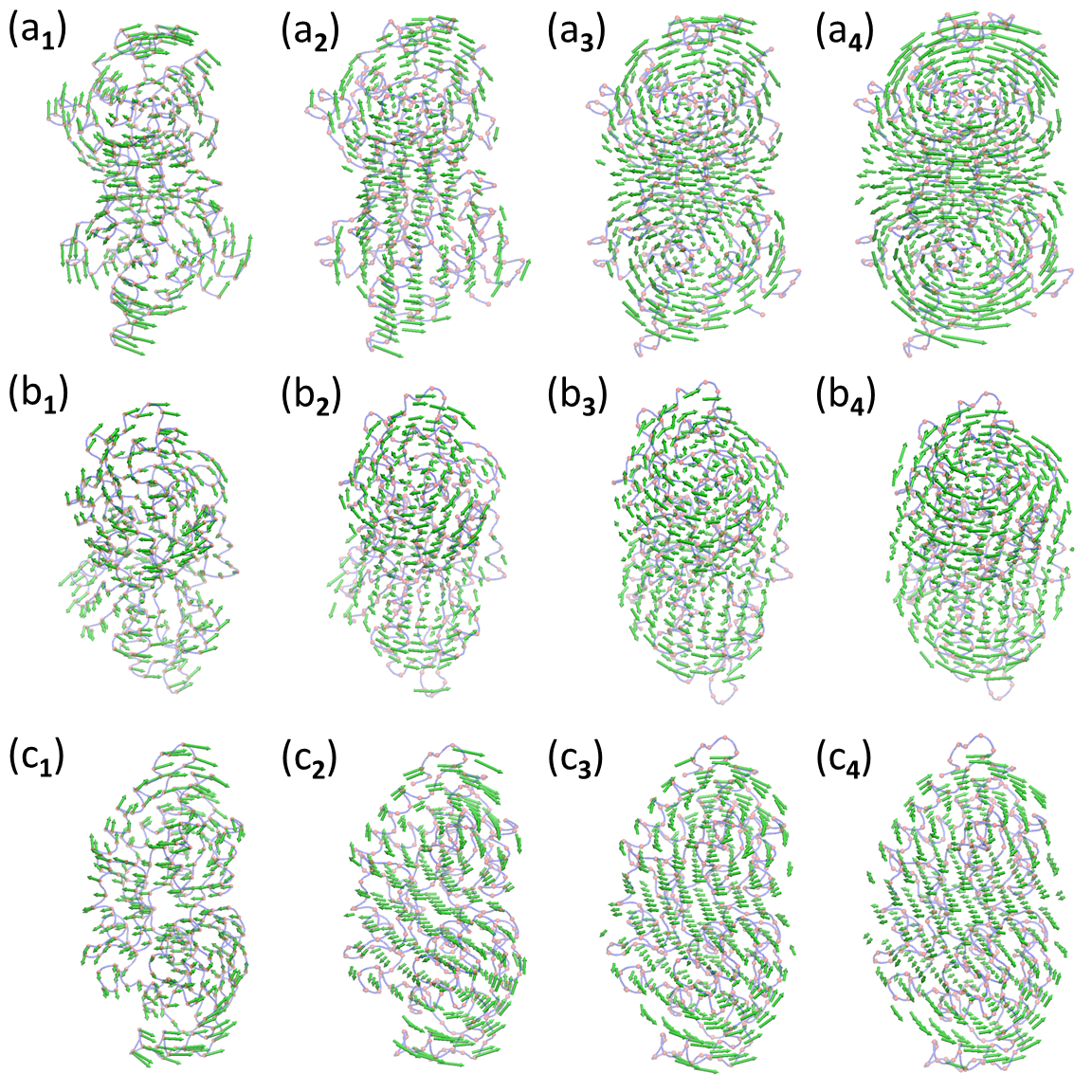}
\end{tabular}
\end{center}
\caption{ The comparison of the three lowest nontrivial eigenmodes from ANM and MVP-ANM for protein 2ABH. The indexes ${\bf a}$, ${\bf b}$ and ${\bf c}$ represent eigenmodes 7 to 9, respectively. Subscript ${\bf 1}$ is for ANM results. Subscripts ${\bf 2}$ to  ${\bf 4}$ are for MVP-ANM results using $\eta=5$\AA, $10$\AA ~ and $15$\AA, respectively. A highly consistent behavior can be observed.
}
\label{fig:2ABH_modes}
\end{figure}

\begin{figure}
\begin{center}
\begin{tabular}{c}
\includegraphics[width=0.6\textwidth]{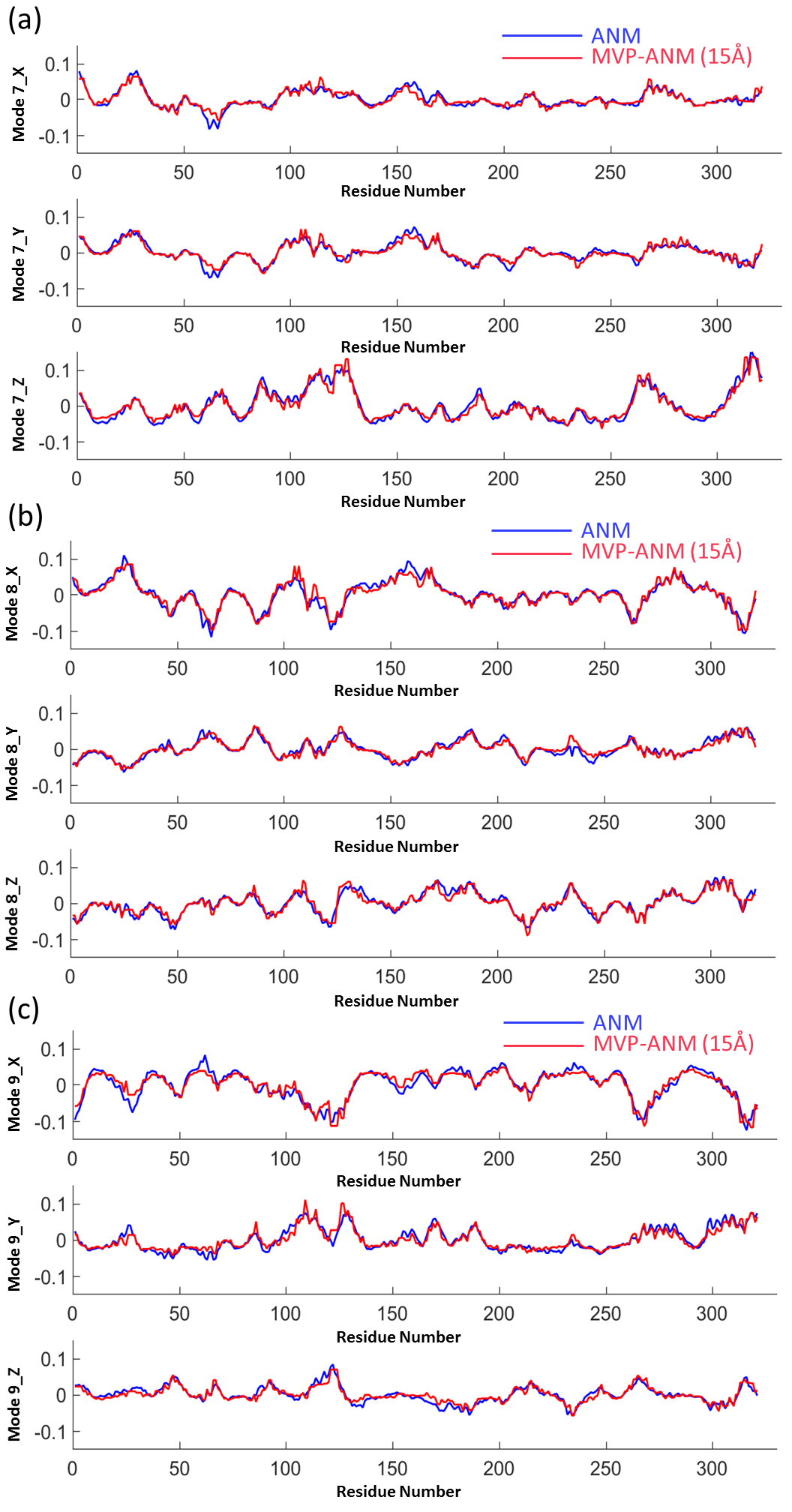}
\end{tabular}
\end{center}
\caption{The detailed comparison of the X, Y and Z components from the three lowest nontrivial eigenmodes from ANM and MVP-ANM for protein 2ABH. The scale value is $15$\AA. $({\bf a})$, $({\bf b})$ and $({\bf c})$ represent eigenmodes 7 to 9, respectively.
}
\label{fig:2ABH_modes_PCC}
\end{figure}

To further explore the performance of my MVP-ANM in normal mode analysis, I compare the three lowest nontrivial eigenmodes from ANM and MVP-ANM from three different scales.
For protein 2ABH, the three eigenmodes predicted from ANM and my MVP-ANM are illustrated in Figure \ref{fig:2ABH_modes}. Here $({\bf a})$, $({\bf b})$ and $({\bf c})$  represent eigenmodes 7 to 9, respectively. Subscript ${\bf 1}$ is for ANM results. Subscripts ${\bf 2}$ to  ${\bf 4}$ are for MVP-ANM results using $\eta=5$\AA, $10$\AA ~ and $15$\AA, respectively. Since a uniformed isovalue is used in MVP-ANM, the shape of a protein, or more specifically the boundary of the protein density region considered in my model, is gradually smoothed as the resolution value becomes larger. And the number of virtual particles increases accordingly. However, there is a high consistence between the predictions from my MVP-ANM and ANM, as can be observed in Figure \ref{fig:2ABH_modes}. My MVP-ANM manages to preserve the eigenmodes from the lowest eigenvalues very well. It is worth mentioning that the indexes of some eigenmodes in my MVP-ANM are renumbered according to ANM eigenmodes. For example, mode 8 from MVP-ANM may correspond to mode 9 in ANM. In that situation, it is renumbered as mode 9.

To have a more quantitative evaluation of the similarities between eigenmodes calculated from the two methods, each eigenmode is decomposed into three individual vectors, representing eigenmode components in X, Y and Z directions, respectively. In MVP-ANM, I only consider eigenmodes from the lowest resolution/scale parameter, i.e., $\sigma=15$\AA. The PCCs for X, Y and Z vectors of Mode 7 between ANM results and MVP-ANM results are 0.914, 0.929 and 0.960, respectively. For Mode 8, the PCCs for X, Y and Z vectors are 0.948, 0.930 and 0.943, respectively. For mode 9, the PCCs for X, Y and Z vectors are 0.936, 0.921 and 0.906, respectively. The corresponding results are illustrated in Figure \ref{fig:2ABH_modes_PCC}. To avoid confusion, the same interpolation from virtual particles onto atomic centers is employed as in the previous cases. 

\begin{figure}
\begin{center}
\begin{tabular}{c}
\includegraphics[width=0.65\textwidth]{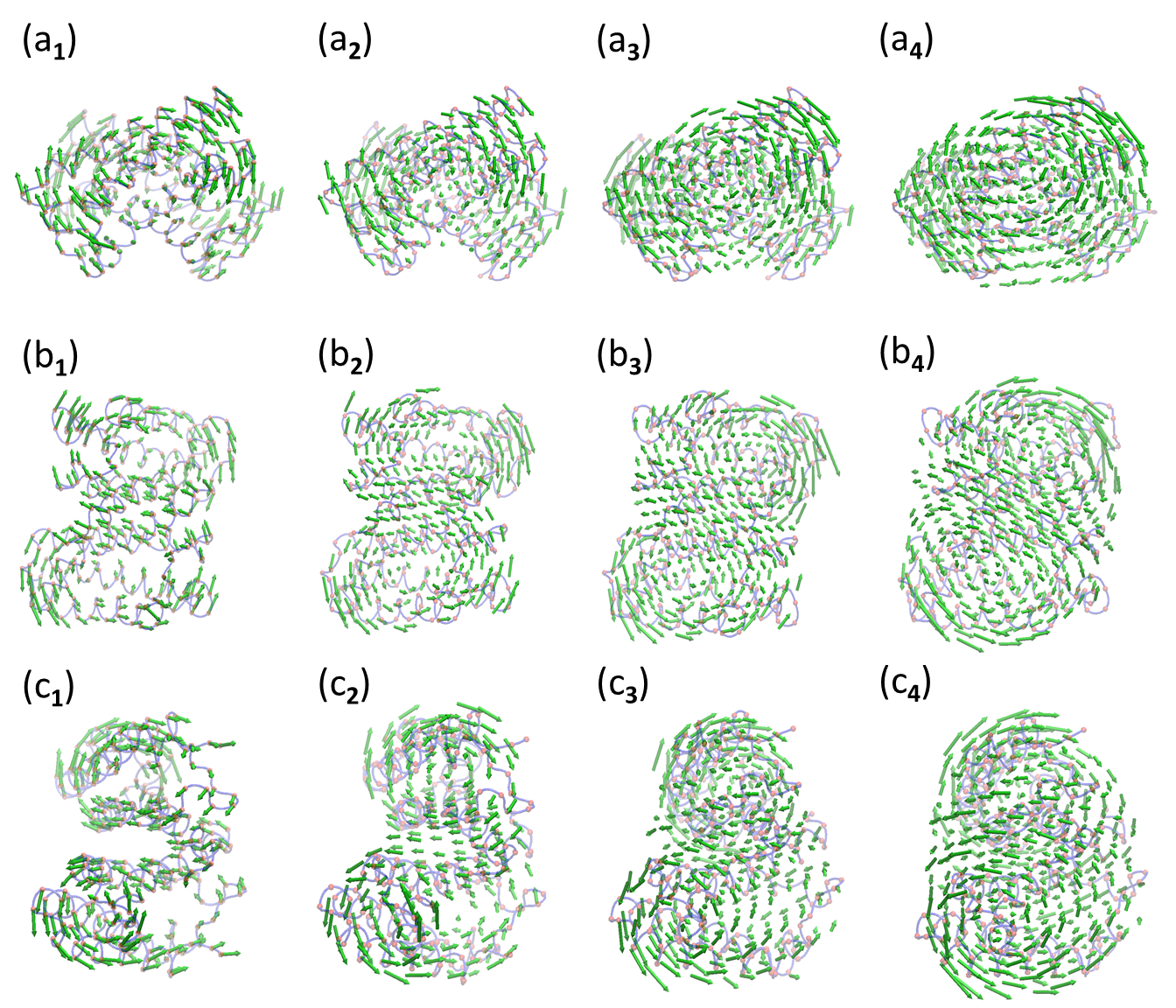}
\end{tabular}
\end{center}
\caption{The comparison of the three lowest nontrivial eigenmodes from ANM and MVP-ANM for protein 2CCY. The indexes ${\bf a}$, ${\bf b}$ and ${\bf c}$ represent eigenmodes 7 to 9, respectively. Subscript ${\bf 1}$ is for ANM results. Subscripts ${\bf 2}$ to  ${\bf 4}$ are for MVP-ANM results using $\eta=5$\AA, $10$\AA ~ and $15$\AA, respectively. A highly consistent behavior can be observed.
}
\label{fig:2CCY_modes}
\end{figure}

\begin{figure}
\begin{center}
\begin{tabular}{c}
\includegraphics[width=0.6\textwidth]{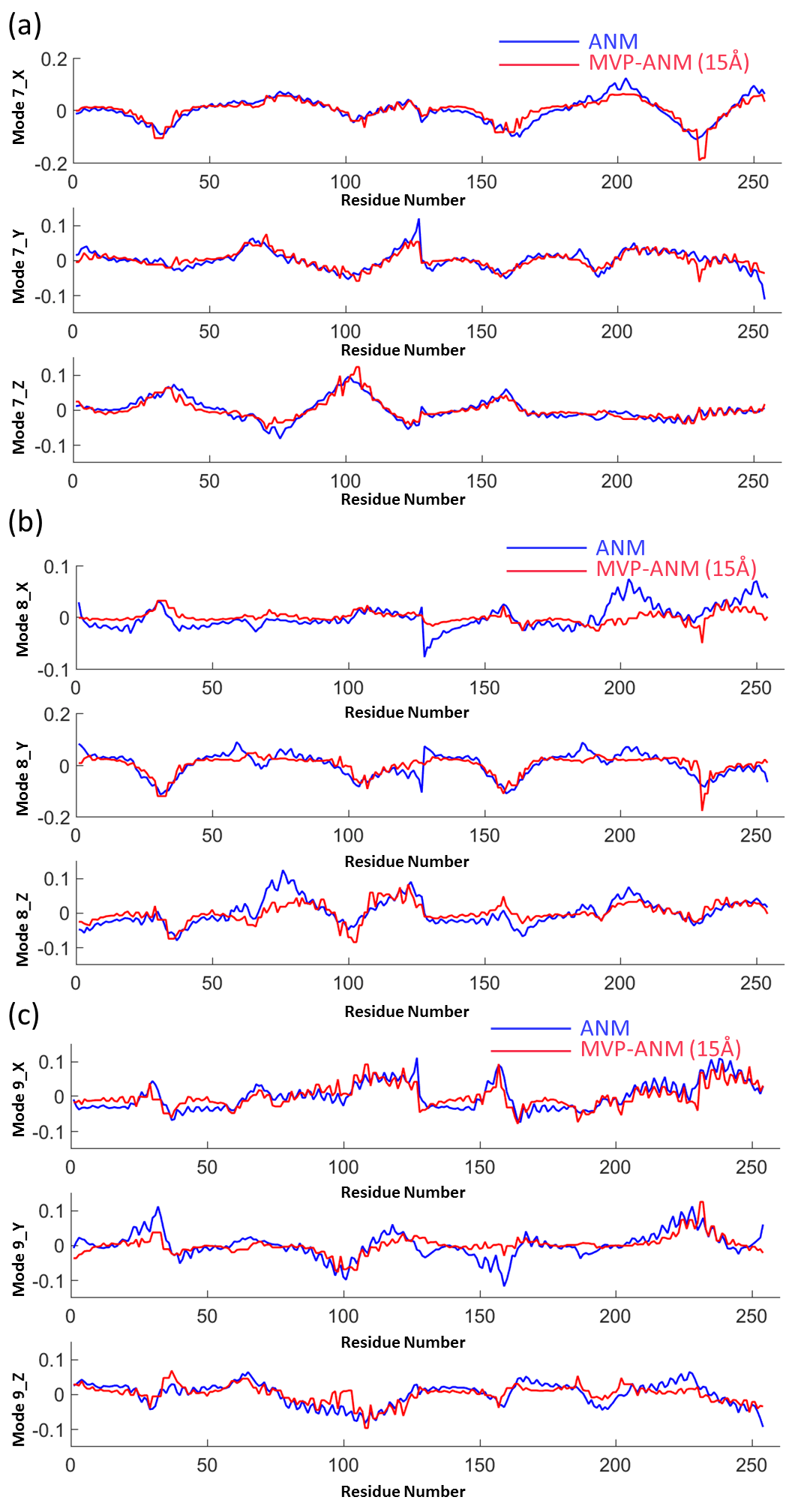}
\end{tabular}
\end{center}
\caption{The detailed comparison of the X, Y and Z components from the three lowest nontrivial eigenmodes from ANM and MVP-ANM for protein 2CCY. The scale value is $15$\AA. $({\bf a})$, $({\bf b})$ and $({\bf c})$ represent eigenmodes 7 to 9, respectively.
}
\label{fig:2CCY_modes_PCC}
\end{figure}

For protein 2CCY, the three eigenmodes predicted from ANM and my MVP-ANM are illustrated in Figure \ref{fig:2CCY_modes}. The subfigures are arranged in the same way with the same notations as Figure \ref{fig:2CCY_modes_PCC}. Indexes of some modes are renumbered for a better visualization and comparison. Again, there is a high consistence between the predictions from my MVP-ANM and ANM. To be more specific, the PCCs for the X, Y and Z vectors of Mode 7 between ANM results and MVP-ANM results are 0.896, 0.834 and 0.910, respectively. For Mode 8, the PCCs are 0.461, 0.799 and 0.746, respectively. For mode 9, the PCCs are 0.827, 0.681 and 0.759, respectively. The results are demonstrated in Figure \ref{fig:2CCY_modes_PCC}.

\begin{figure}
\begin{center}
\begin{tabular}{c}
\includegraphics[width=0.4\textwidth]{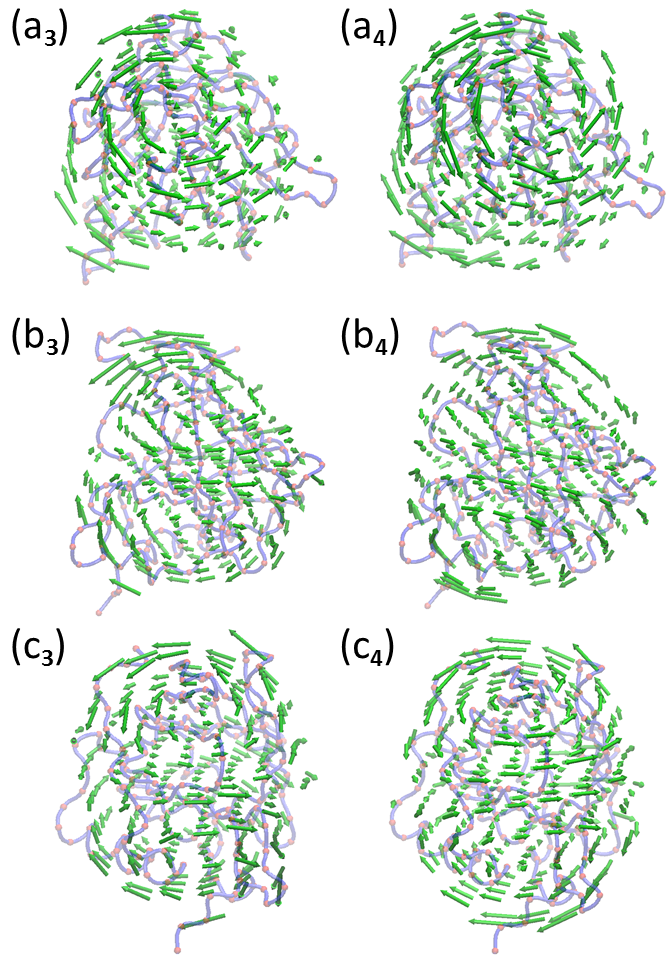}
\end{tabular}
\end{center}
\caption{The comparison of the three lowest nontrivial eigenmodes from MVP-ANM for protein 1AQB. The indexes $({\bf a})$, $({\bf b})$ and $({\bf c})$ represent eigenmodes 7 to 9, respectively. Subscripts ${\bf 3}$ to  ${\bf 4}$ are for MVP-ANM results using $\eta=10$\AA ~and $15$\AA, respectively.
}
\label{fig:1AQB_modes}
\end{figure}

For protein 1AQB, its eigenmodes from MVP-ANM are illustrated in Figure \ref{fig:1AQB_modes}. The subfigures are arranged in the same way with the same notations.  However, I only depict the results from two resolutions, i.e., $\sigma=10$\AA~ and $15$\AA,~ because the eigenmodes from ANM and MVP-ANM with $\sigma=5$\AA~ have unproportionately large local eigenvectors at either extruding loop regions or ends of chains. The magnitude of these a few local eigenvectors is so large that it overshadows the eigenmode information of all other atoms. However, when the resolution value is large enough, the large fluctuation in these local eigenvectors are attenuated and global motions begin to emerge. There is a clear consistence of the eigenmodes from resolution $\sigma=10$\AA~ and $15$\AA~ models.

From the above three cases, it can be seen that my MVP-ANM is able to preserve the basic collective motions very well even at very low resolutions. This confirms that the collective motions are primarily determined by the global shapes and mass distributions and insensitive to local structure variations.


\paragraph{Shape influence}
\begin{figure}
\begin{center}
\begin{tabular}{c}
\includegraphics[width=0.95\textwidth]{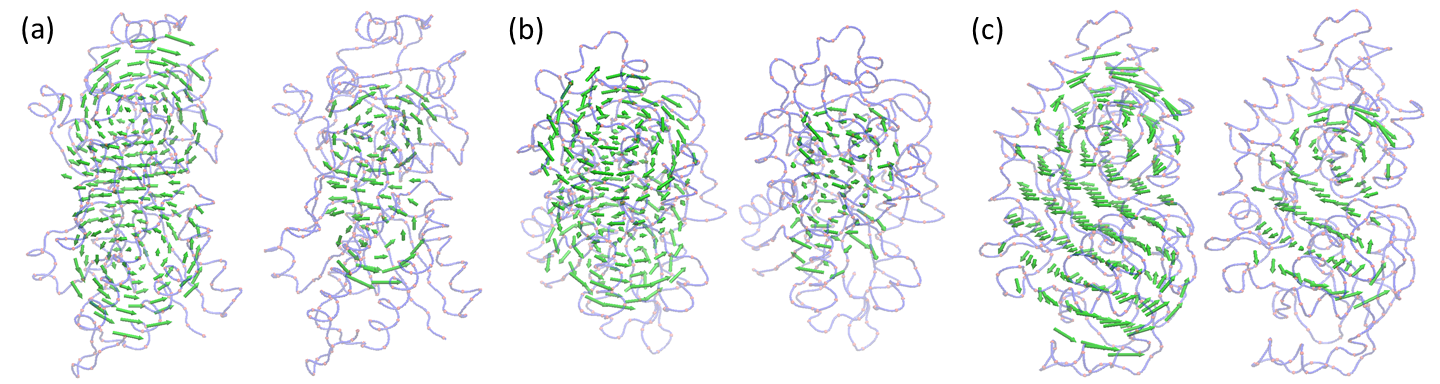}
\end{tabular}
\end{center}
\caption{ The comparison of the three lowest nontrivial eigenmodes from MVP-ANM using different isovalues. The value resolution parameter is $10$\AA.~$({\bf a})$, $({\bf b})$ and $({\bf c})$ represent eigenmodes 7 to 9, respectively. Two different isovalues, i.e., $0.6$ and $0.8$, are used with results demonstrated on the left and right hand side, respectively, in each eigenmode.
}
\label{fig:2ABH_iso}
\end{figure}

It has been well recognized that the general shape of a protein has played a critical role in its collective motions. However, a well-defined shape or biomolecular surface is not always available for a biomolecular density data. Even though sometime an isovalue is suggested, it is mainly for the visualization purpose. The influence of different isovalues on the collective motions remains unclear.

In this part, I compare the three lowest nontrivial eigenmodes from MVP-ANM with different computational domains. These domains are generated by a series of isovalues. It is found that the eigenmodes from my MVP-ANM share a very consistent pattern over a wide range of isovalues.

To have a better understand of this, I use protein 2ABH as an example. The density model is generated in the same way as in the previous cases. Here I only consider the density profile generated by resolution $\eta=10$\AA.~ The detailed results are demonstrated in Figure \ref{fig:2ABH_iso}. Here $({\bf a})$, $({\bf b})$ and $({\bf c})$  represent eigenmodes 7 to 9, respectively. Two different isovalues, i.e., $0.6$ and $0.8$, are chosen in MVP-ANM. The results from the two isovalues are demonstrated on the left and right hand side, respectively, in each eigenmode. The total numbers of virtual particles from these two isovalues differ greatly. To be more specific, there are 321 $C_{\alpha}$ atoms in 2ABH. When isovalue equals to $0.4$, there are totally 372 virtual particles. This number drops to 195 when isovalue equals $0.6$ and further plummets to 77 if isovalue $0.8$ is used. However, the collective motions of the protein are still well preserved in my model. As can be seen from Figure \ref{fig:2ABH_iso}, the demonstrated modes 7 to 9 are very consistent with the results in Figure \ref{fig:2ABH_modes}. Even when the virtual particles reduce to less than a quarter of the total atoms (when isovalue is $0.6$), the general pattern of the collective motions still remains the same.

From the above analysis, it can be seen that the MVP-ANM can deliver very consistent collective motions for a wide range of isovalues. This is largely due to the reason that when a coarse resolution is used, the topology of the general shapes is very consistent. Therefore the pattern of the collective motions remains the same. More importantly, this means that in virtual particle generation, the MVP-ANM has certain flexibility in the isovalue selection.

\section{Application}\label{sec:application}
In this section, my MVP-ANM method is used in the collective motion analysis of large-sized biomolecular structures. Two poliovirus structures with different data types are considered. The first one, PDB 1XYR, has detailed atomic information. The second structure, EMD5122, is represented in the density distribution profile. It should be noticed that the focus of this part is not to list all detailed collective motions of the structure, but to validate the MVP-ANM and explore its computational efficiency. Therefore, only three types of lowest nontrivial eigenmodes are considered.

\begin{figure}
\begin{center}
\begin{tabular}{c}
\includegraphics[width=0.9\textwidth]{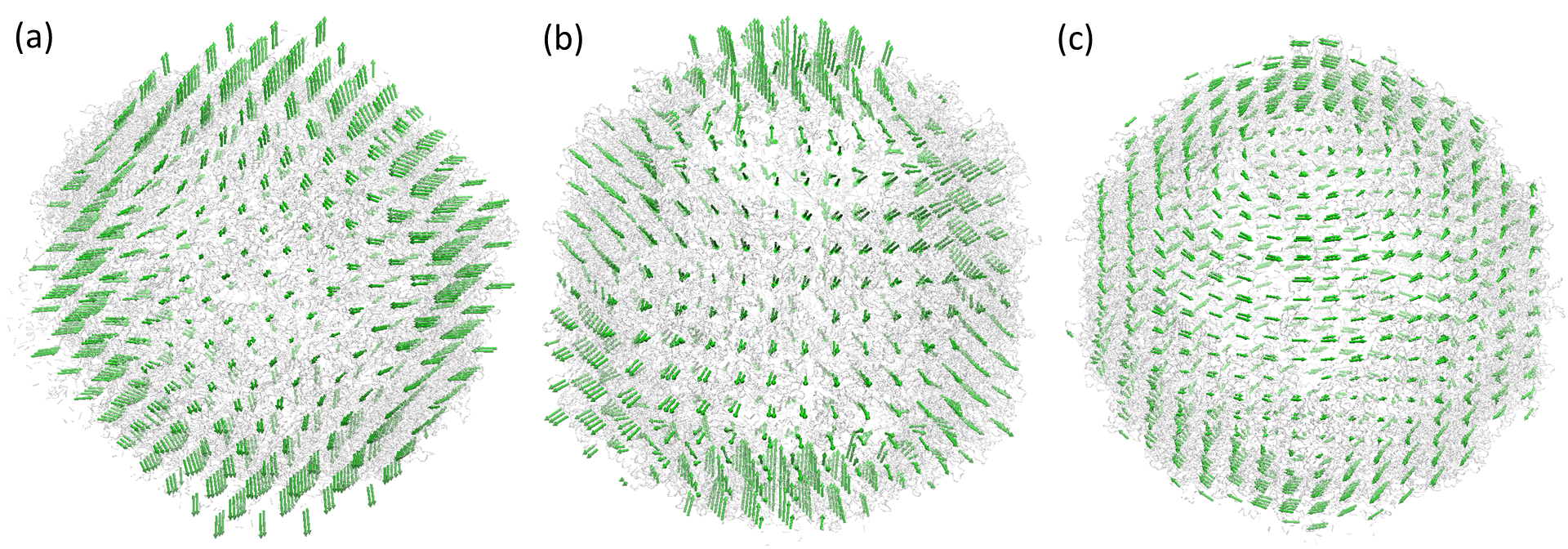}
\end{tabular}
\end{center}
\caption{ The illustration of the three types of lowest nontrivial eigenmodes of poliovirus structure with ID 1XYR.
}
\label{fig:1xyr}
\end{figure}

\begin{figure}
\begin{center}
\begin{tabular}{c}
\includegraphics[width=0.9\textwidth]{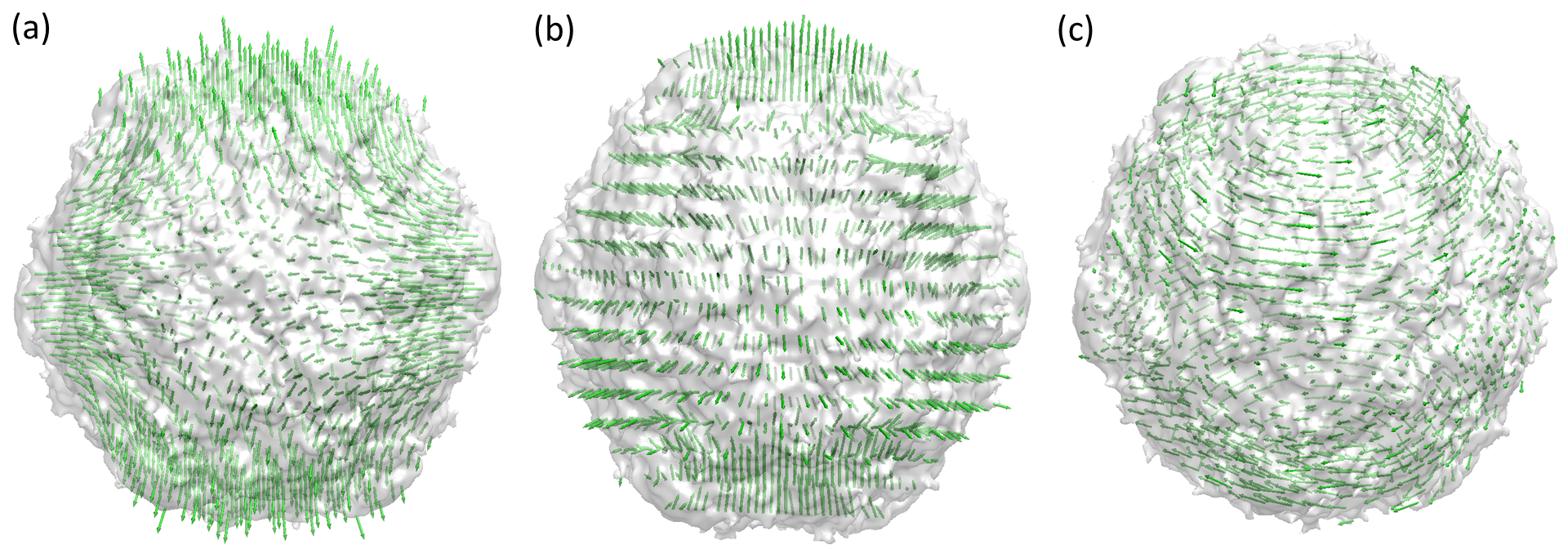}
\end{tabular}
\end{center}
\caption{The illustration of the three types of lowest nontrivial eigenmodes of poliovirus structure with ID  EMD5122.
}
\label{fig:emd5122}
\end{figure}

\paragraph{Case 1: Structure with atomic details}

Protein 1XYR is a poliovirus 135S cell entry intermediate. To analyze its collective motions with MVP-ANM, a density data is constructed first. Similar to the previous cases, the generalized Gaussian model in Eq. (\ref{eq:couple_matrix1}) with $\kappa=2$ and $\eta=4$\AA~ is used. The density data is represented in a volumetric data with grid spacing of 4\AA. Virtual particles are generated by selecting all voxels with normalized density values larger than or equal to 0.13. If we choose each voxel as a virtual particle, the number of voxels considered is too large thus the model is computationally intractable. Therefore, a coarse-graining process is done by combining the adjacent five voxels in each direction together into a single virtual particle. To be more specific, each new virtual particle includes totally 125 voxels from the original density data. The scale parameter in the spring constant is chosen as the size of the voxel, which is 20 \AA.~ With this setting, there are totally 1008 virtual particles in our MVP-ANM model, which dramatically reduce the computational cost.

The results of the three types of lowest nontrivial eigenmodes are illustrated in Figure \ref{fig:1xyr}. Due to the symmetry properties of the structure. Each type of normal modes includes several eigenvectors with different orientations. These three types of motions are widely observed in poliovirus virus\cite{van:2005normal}. These types of motions are highly related to the icosahedron structure of the virus.

\paragraph{Case 1: Structure represented in density profile}

In the second example, the poliovirus EMD5122 is considered. This is a RNA-releasing poliovirus intermediate generated from Single particle reconstruction. This data has 201*201*201 voxels. The voxel spacing is 2.322\AA. The threshold value is chosen as 10 to construct the proper virtual particle model. That is to say, all volxels with density value larger than 10 is treated as virtual particles. This, again, will result in a huge number of virtual particles beyond computational limit. Therefore, a coarse-graining process is done by combining the adjacent 8 voxels in each direction together into a single virtual particle. To be more specific, each new virtual particle includes totally 512 voxels from the original data. With this setting, there are only 1267 virtual particles in the MVP-ANM, which again has reduced the computational complexity tremendously.

The results of the eigenmodes are depicted in Figure \ref{fig:emd5122}. I do not list all the possible eigenmodes. Instead three types of lowest nontrivial eigenmodes are considered. By the comparison of the results in Figure \ref{fig:1xyr} and Figure \ref{fig:emd5122}, it can be observed that two structures share almost the identical types of collective motions. This is not a surprise, as both have a rather regular icosahedron symmetry structure. Different scales of virtual particle model are also employed, there is a great consistence of these types of collective motions. The results again confirm that biomolecular collective motions are directly related to their general shapes.

\section{Conclusion remarks}\label{sec:conclusion}
A multiscale virtual particle based elastic network model (MVP-ENM) is proposed for biomolecular normal mode analysis. Similar to the previous VP-aFRI model, MVP-ENM is designed for biomolecular density data analysis.  The basic process in the MVP-ENM is as following. Firstly, a molecular surface is extracted by using a suitable isovalue. Essentially, the surface is treated as the boundary of our computational domain. Secondly, the inside region/regions of the biomolecule, i.e., the domain/domains enclosed by this surface, are discretized. Various methods, including the Cartesian grid in the finite difference method, the tetrahedron mesh in the finite element method, the Voronoi cell in the tessalation method, etc, can all be employed in the domain discretization. The resulting elements are virtual particles. Essentially, this works as a coarse-graining (CG) of biomolecular structures, so that a delicate balance between biomolecular geometric representation and computational cost can be achieved. To form ``connections" between these virtual particles, a new harmonic potential function, which considers the influence from both mass distributions and distance relations, is adopted between any two virtual particles.

Two independent models, i.e., multiscale virtual particle based Gaussian network model (MVP-GNM) and multiscale virtual particle based anisotropic network model (MVP-ANM), are included in MVP-ENM. It is found that even with a related coarse grid and a low resolution, the MVP-GNM is able to predict the Debye-Waller factors with considerable good accuracy. The mismatch is predominantly from higher fluctuation regions. MVP-ANM shares similar behavior as MVP-GNM in B-factor prediction. More interestingly, MVP-ANM delivers a very consistent low-frequency eigenmodes in various scales. This demonstrates the great potential of MVP-ANM in the deformation analysis of low resolution data. It is worth mentioning that the MVP-ENM can used be applied to biomolecular data, represented both in density distribution and atomic coordinates. It is a highly efficient tool for analyzing large-sized biomolecules.

\section*{Acknowledgments}
This work was supported in part by Nanyang Technological University Startup Grant M4081842.110 and Singapore Ministry of Education Academic Research fund Tier 1 M401110000.

\vspace{0.6cm}
\clearpage


\end{document}